\documentclass[10pt,conference]{IEEEtran}

\IEEEoverridecommandlockouts

\usepackage{amsfonts}
\usepackage[dvips]{graphicx}
\usepackage{times}
\usepackage{cite}
\usepackage{amsmath}
\usepackage{cases}
\usepackage{array}
\usepackage{dsfont}
\usepackage{amssymb}

\usepackage{stfloats}
\usepackage{slashbox}
\usepackage{graphicx}
\usepackage{footnote}
\usepackage{booktabs}
\usepackage{array}
\usepackage{bm}
\usepackage{epstopdf}
\usepackage{epsfig}
\usepackage[labelformat=simple]{subcaption}

\begin{document}

\title{Cooperative Tx/Rx Caching in Interference Channels: A Storage-Latency Tradeoff Study}
\author{\IEEEauthorblockN{Fan Xu, Kangqi Liu and Meixia Tao}
\IEEEauthorblockA{Dept. of Electronic Engineering, Shanghai Jiao Tong University,  Shanghai, China\\
\thanks{This work is supported by the NSF of China under grants 61571299, 61322102,
and 61329101.}
Emails: xxiaof@sjtu.edu.cn, k.liu.cn@ieee.org, mxtao@sjtu.edu.cn}
}
\maketitle

\begin{abstract}
This paper studies the storage-latency tradeoff in the $3\times3$ wireless interference network with caches equipped at all transmitters and receivers. The tradeoff is characterized by the so-called \textit{fractional delivery time} (FDT) at given normalized transmitter and receiver cache sizes. We first propose a generic cooperative transmitter/receiver caching strategy with adjustable file splitting ratios. Based on this caching strategy, we then design the delivery phase carefully to turn the considered interference channel opportunistically into broadcast channel, multicast channel, X channel, or a hybrid form of these channels. After that, we obtain an achievable upper bound of the minimum FDT by solving a linear programming problem of the file splitting ratios. The achievable FDT is a convex and piece-wise linear decreasing function of the cache sizes. Receiver local caching gain, coded multicasting gain, and transmitter cooperation gain (interference alignment and interference neutralization) are leveraged in different cache size regions.
\end{abstract}


\section{Introduction}
Mobile data traffic has been shifting from connection-centric services, such as voice, e-mails and web browsing, to emerging content-centric services, such as video streaming, push media, application download/updates, and mobile TV. The contents in these services are typically produced well ahead of transmission and can be requested by multiple users, although at possibly different times. This allows us to cache the contents at the edge of wireless networks, e.g., base stations and user devices, and hence to reduce user access latency and alleviate wireless traffic. A fundamental question in wireless cache networks is what and how much gain can be leveraged through caching.

Caching in a shared link with one server and multiple cache-enabled users is first studied by Maddah-Ali and Niesen in \cite{fundamentallimits}. It is shown that caching at user ends brings not only local caching gain but also global caching gain. The latter is achieved by a carefully designed cache placement and coded delivery strategy, which can create multicast chances for content delivery even if users demand different files. The idea is then extended to the decentralized coded caching in a large network in \cite{decentralized}. In \cite{cache&csit}, the authors considered the wireless broadcast channel with imperfect channel state information at the transmitter (CSIT) and showed that the gain of coded multicasting scheme can offset the loss due to the imperfect CSIT.

The authors in \cite{upperbound} studied the transmitter cache strategy in the cache-aided interference channel. It is shown that splitting contents into different parts and caching each part in different transmitters can turn the interference channel into broadcast channel, X channel, or hybrid channel and hence increase the system throughput via interference management. The authors in \cite{lowerbound} presented a lower bound of delivery latency in a general interference network with transmitter cache and showed that the scheme in \cite{upperbound} is optimal in certain region of cache size.

The above literature reveals that caching at the receiver side can bring coded multicasting gain, and that caching at the transmitter side can induce transmitter cooperation for interference management. It is thus of great interests to investigate the impact of caching at both transmitter and receiver sides.

\begin{figure}[!tbp]
\begin{centering}
\includegraphics[scale=0.19]{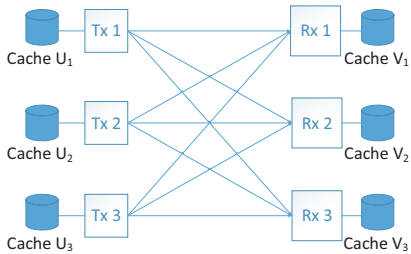}
\vspace{-0.2cm}
\caption{3$\times$3 Interference channel with cache at Tx/Rx sides.}\label{Fig model}
\vspace{-0.5cm}
\end{centering}
\end{figure}

In this paper, we aim to study the fundamental limits of caching in the $3\times3$ interference network with caches equipped at all transmitters and receivers as shown in Fig.~\ref{Fig model}. We adopt the storage-latency tradeoff originally proposed in \cite{lowerbound} to characterize the fundamental limits. In specific, we measure the performance by the \textit{fractional delivery time} (FDT) as a function of the normalized receiver and transmitter cache sizes. To analyze the minimum FDT, we propose a generic file splitting and caching strategy with adjustable file splitting ratios. Based on this strategy, we then design the delivery phase carefully so that the network topology can be opportunistically changed to broadcast channel, multicast channel, X channel, or a hybrid form of these channels. We then obtain an achievable upper bound of the minimum FDT by optimizing the file splitting ratios. The obtained FDT is a convex and piece-wise linear decreasing function of the transmitter and receiver cache sizes. Our result shows that coded multicasting gain should be exploited as much as we can when the cache sizes are very limited. It also shows that transmitter cooperation gain can only be exploited when the transmitter cache size exceeds a certain threshold dependent on the receiver cache size. Note that an independent work on the similar problem is studied in \cite{bothcache}. We shall discuss the differences with \cite{bothcache} later.

Notations: $(\cdot)^T$ denotes the transpose. $[K]$ denotes set $\{1,2,\cdots,K\}$. $\lfloor x\rfloor$ denotes the largest integer no greater than $x$. $(x_j)^K_{j=1}$ denotes vector $(x_1,x_2,\cdots,x_K)^T$. $\mathcal{CN}(m,\sigma^2)$ denotes the complex Gaussian distribution with mean of $m$ and variance of $\sigma$.

\section{System Model and Definitions}
Consider the 3$\times$3 cache-aided interference channel shown in Fig.~\ref{Fig model}. Each node is assumed to have single antenna. The communication link between each transmitter and each receiver experiences channel fading and is corrupted with additive white Gaussian noise. The communication at each time slot $t$ over this network is modeled by
\begin{align}
Y_j(t)=\sum_{p=1}^{3}h_{jp}(t)X_p(t)+Z_j(t),j=1,2,3,\notag
\end{align}
where $Y_j(t)\in \mathbb{C}$ denotes the received signal at receiver $j$, $X_p(t)\in \mathbb{C}$ denotes the transmitted signal at transmitter $p$, $h_{jp}(t)\in \mathbb{C}$ denotes the channel coefficient from transmitter $p$ to receiver $j$, and $Z_j(t)$ denotes the noise at receiver $j$ distributed as $\mathcal{CN}(0,1)$.

Consider a database consisting of $L$ files ($L>> 3$), denoted by $\{W_1,W_2,\cdots,W_L\}$. Each file is chosen independently and uniformly from $[2^F]=\{1,2,\cdots,2^F\}$ randomly, where $F$ is the file size in bits. Each transmitter has a local cache able to store $M_TF$ bits and each receiver has a local cache able to store $M_RF$ bits. The \textit{normalized cache sizes} at each transmitter and receiver are defined, respectively, as
\begin{eqnarray}
\mu_T\triangleq\frac{M_T}{L},\qquad\mu_R\triangleq\frac{M_R}{L}.\notag
\end{eqnarray}

The network operates in two phases, \textit{cache placement phase} and \textit{content delivery phase}. During the cache placement phase, each transmitter $p$ designs a caching function
\begin{eqnarray}
\phi_p:[2^F]^L\to[2^{\lfloor FM_T\rfloor}],\notag
\end{eqnarray}
mapping the $L$ files in the database to its local cached content $U_p\triangleq\phi_p(W_1,W_2,\cdots,W_L)$. Each receiver $j$ also designs a caching function
\begin{eqnarray}
\psi_j:[2^F]^L\to[2^{\lfloor FM_R\rfloor}],\notag
\end{eqnarray}
mapping the $L$ files to its local cached content $V_j\triangleq\psi_j(W_1,W_2,\cdots,W_L)$. The caching functions $\{\phi_p,\psi_j\}$ are assumed to be known globally at all nodes. In the delivery phase, each receiver $j$ requests a file $W_{d_j}$ from the database. We denote ${\bf d}\triangleq(d_j)^{3}_{j=1}\in[L]^{3}$ as the demand vector. Each transmitter $p$ has an encoding function
\begin{eqnarray}
\Lambda_p:[2^{\lfloor FM_T\rfloor}]\times[L]^{3}\times\mathbb{C}^{3\times 3}\to\mathbb{C}^T.\notag
\end{eqnarray}
Transmitter $p$ uses $\Lambda_p$ to map its cached content $U_p$, receiver demands ${\bf d}$ and channel realization $\mathbf{H}$ to the signal $(X_p[t])^T_{t=1}\triangleq\Lambda_p(U_p,{\bf d},\mathbf{H})$, where $T$ is the block length of the code. Note that $T$ may depend on the receiver demand ${\bf d}$ and channel realization $\mathbf{H}$, and thus can also be denoted as $T^{\mathbf{d},\mathbf{H}}$ (with a slight abuse of notation, we will use $T$ again to denote the average worst-case delivery time in Definition 1). Each codeword $(X_p[t])_{t=1}^T$ has an average transmit power constraint $P$. Each receiver $j$ has a decoding function
\begin{eqnarray}
\Gamma_j:[2^{\lfloor FM_R\rfloor}]\times\mathbb{C}^T\times\mathbb{C}^{3\times 3}\times[L]^{3}\to[2^F].\notag
\end{eqnarray}
Upon receiving $(Y_j[t])^T_{t=1}$, each receiver $j$ uses $\Gamma_j$ to decode $\hat{W}_j\triangleq\Gamma_j(V_j,(Y_j[t])^T_{t=1},\mathbf{H},{\bf d})$ of its desired file $W_{d_j}$ using its cached content $V_j$ as side information. The worst-case error probability is
\begin{eqnarray}
P_\epsilon=\max\limits_{{\bf d}\in[L]^{3}}\max\limits_{j\in[3]}\mathbb{P}(\hat{W}_j\ne W_{d_j}).\notag
\end{eqnarray}
The given caching and coding scheme $\{\phi_p,\psi_j,\Lambda_p,\Gamma_j\}$ is said to be feasible if $P_\epsilon\to 0$ when $F\to\infty$.


In this work, we adopt the following latency-oriented performance metrics originally proposed in \cite{lowerbound}.

\textbf{Definition 1}: The delivery time (DT) for a given feasible caching and coding scheme is defined as
\begin{eqnarray}
T\triangleq\lim\limits_{P\to\infty}\lim\limits_{F\to\infty}\max\limits_{{\bf d}}\mathbb{E}_\mathbf{H}(T^{{\bf d},\mathbf{H}}).
\end{eqnarray}

\textbf{Definition 2}: The \textit{fractional delivery time} (FDT) for a given feasible caching and coding scheme is defined as
\begin{eqnarray}
\tau(\mu_R,\mu_T)\triangleq\lim_{P\to\infty}\lim_{F\to\infty}\sup\frac{\max\limits_{{\bf d}}\mathbb{E}_\mathbf{H}(T^{{\bf d},\mathbf{H}})}{N_RF\cdot1/\log P},\notag
\end{eqnarray}
where $N_R=3$ is the number of content requesters. Moreover, the minimum FDT at given normalized cache sizes $\mu_T$ and $\mu_R$ is defined as
\begin{eqnarray}
\tau^*(\mu_R,\mu_T)=\inf\{\tau(\mu_R,\mu_T):\tau(\mu_R,\mu_T)\rm \;is\;achievable\}.\notag
\end{eqnarray}

The above performance metrics are defined in the asymptomatic sense when $P \to \infty$ and $F\to \infty$. It is clear that the FDT and DT are related by $\tau=\frac{T\log P}{3F}$. The FDT $\tau(\mu_R,\mu_T)$ can be regarded as the relative time with respect to delivering the total $3F$ requested bits in an  interference-free baseline system with transmission rate $\log P$ in the high SNR region.

\textbf{Remark 1}: Our definition of FDT $\tau$ is slightly different from the \textit{normalized delivery time} (NDT) $\delta$ in \cite{lowerbound} in that our FDT is further normalized by the number of receivers. That is, $\tau=\delta/ 3$. With such normalization, the FDT is defined for the total $3F$ bits requested in the network rather than the $F$ bits requested by a single receiver as in \cite{lowerbound}. As a result, the range of FDT is $0\le\tau\le1$, which is truly normalized.

\textbf{Remark 2}: Compared to the ``load'' $R$ defined for the shared link in \cite{fundamentallimits}, the FDT can be expressed as $\tau=\frac{R}{3\cdot \textrm{DoF}}$, where DoF is the sum DoF of the considered channel. Comparing to the standard DoF adopted for interference channel with transmitter cache only in \cite{upperbound}, we have $\tau(\mu_R=0, \mu_T) = \frac{1}{\textrm{DoF}}$. As a result, the FDT evaluates the delivery time of the actual \textit{load} at a transmission rate specified by \textit{DoF} of the given channel, and hence is particularly suitable to characterize the performance of the wireless network with both transmitter and receiver caches.

\textbf{Remark 3} (Feasible domain of FDT): The FDT introduced above is able to measure the fundamental tradeoff between the cache storage and content delivery latency. However, not all normalized cache sizes are feasible. Given fixed $L$ and $M_T$, all the transmitters together can store at most $3M_TF$ bits of files, which leaves $LF-3M_TF$ bits of files to be stored in all receivers. Thus we must have $M_RF\ge LF-3M_TF$. This gives the feasible region for the normalized cache sizes as:
\begin{align}
\left\{
\begin{array}{ll}0\le\mu_R,\mu_T\le1\\ \mu_R+3\mu_T\ge1
\end{array}.\label{eqn boundary}
\right.
\end{align}
Throughout this paper, we study the FDT only in the feasible domain \eqref{eqn boundary}.
\begin{figure}[tbp]
\begin{centering}
\includegraphics[scale=0.35]{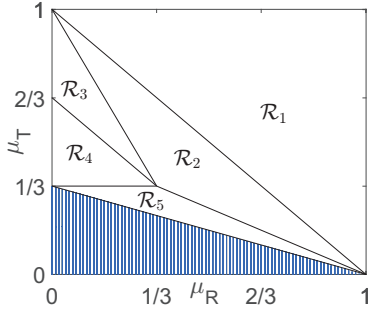}
\vspace{-0.2cm}
\caption{Feasible domain of FDT (divided into 5 regions).}\label{Fig region}
\vspace{-0.6cm}
\end{centering}
\end{figure}
\section{Main Results}
In this section, we present an achievable upper bound of the minimum FDT $\tau^*(\mu_R,\mu_T)$. The proof will be given in the next two sections.

\textbf{Theorem 1}: For the $3\times3$ cache-aided interference channel, the minimum FDT is upper bounded by
\begin{align}
\tau^*(\mu_R,\mu_T)\le
\left\{
\begin{array}{ll}
\frac{1}{3}-\frac{\mu_R}{3}, & (\mu_R,\mu_T)\in \mathcal{R}_1\\
\frac{4}{9}-\frac{4\mu_R}{9}-\frac{\mu_T}{9}, & (\mu_R,\mu_T)\in \mathcal{R}_2\\
\frac{1}{2}-\frac{5}{9}\mu_R-\frac{1}{6}\mu_T, & (\mu_R,\mu_T)\in \mathcal{R}_3\\
\frac{13}{18}-\frac{8}{9}\mu_R-\frac{1}{2}\mu_T, & (\mu_R,\mu_T)\in \mathcal{R}_4\\
\frac{8}{9}-\frac{8}{9}\mu_R-\mu_T, & (\mu_R,\mu_T)\in \mathcal{R}_5
\end{array},\notag
\right.
\end{align}
where $\{\mathcal{R}_i\}^5_{i=1}$ are given below and sketched in Fig.~\ref{Fig region}.
\begin{align}
\left\{
\begin{array}{ll}
&\mathcal{R}_1=\{(\mu_R,\mu_T): \mu_R+\mu_T\ge1, \mu_R\le1, \mu_T\le1\}\\
&\mathcal{R}_2=\{(\mu_R,\mu_T): \mu_R+\mu_T<1, 2\mu_R+\mu_T\ge1, \\
&\qquad\qquad\qquad\qquad\mu_R+2\mu_T>1\}\\
&\mathcal{R}_3=\{(\mu_R,\mu_T): \mu_R+\mu_T\ge\frac{2}{3}, 2\mu_R+\mu_T<1, \\
&\qquad\qquad\qquad\qquad\mu_R\ge0\}\\
&\mathcal{R}_4=\{(\mu_R,\mu_T): \mu_R+\mu_T<\frac{2}{3}, \mu_R\ge0, \mu_T>\frac{1}{3}\}\\
&\mathcal{R}_5=\{(\mu_R,\mu_T): \mu_T\le\frac{1}{3}, \mu_R+2\mu_T\le1, \\
&\qquad\qquad\qquad\qquad\mu_R+3\mu_T\ge1\}\notag
\end{array}.
\right.
\end{align}
%


The above theorem shows that the achievable FDT is a convex and piecewise linear decreasing function of $\mu_R$ and $\mu_T$. It captures an achievable tradeoff between the cache storage and the delivery latency. In the special case when $\mu_R=0$ (transmitters cache only), the results reduce to
\begin{align}
\tau^*(0,\mu_T)\le
\left\{
\begin{array}{ll}
13/18-\mu_T/2, & 1/3\le\mu_T\le2/3\\
1/2-\mu_T/6, &2/3<\mu_T\le1
\end{array},
\right.\notag
\end{align}
which is the same as the upper bound of 1/DoF in \cite{upperbound}.

When $\mu_T=1$, each transmitter can cache all the files and hence the network can be viewed as a virtual broadcast channel as in \cite{fundamentallimits} except that the transmitter has 3 distributed antennas. Thus, we can achieve  FDT $\tau=\frac{1}{3}-\frac{\mu_R}{3}$ here. Comparing to the result in \cite{fundamentallimits}, i.e., $\tau=\frac{1-\mu_R}{1+3\mu_R}$ at $\mu_R=\{0,\frac{1}{3},\frac{2}{3},1\}$, we can see that our FDT is better when $0\le\mu_R<\frac{2}{3}$ and they are same when $\frac{2}{3}\le\mu_R\le1$. The performance improvement is due to transmitter cooperation gain.



\section{Achievable Caching scheme}
\subsection{File Splitting and Placement}
Given fixed $\mu_R$ and $\mu_T$, the content placement can be established as follows.

In this work, we treat all the files equally without taking file popularity into account. Thus, each file will be split and cached in the same manner. Without loss of generality, we focus on the splitting and caching of file $W_i$ for any $1\le i\le L$. Since each bit of the file is either cached or not cached in every node, there are $2^6=64$ possible cache states for each bit. However, note that every bit of the file must be cached in at least one node. In addition, every bit that is not cached simultaneously in all receivers must be cached in at least one transmitter. This is because we do not allow receiver cooperation and all the messages must be sent from the transmitters. As such, the total number of feasible cache states for each bit is given by $64-1-\binom{3}{1}-\binom{3}{2}=57$. Now with possibly different lengths, we can partition each $W_i$ into 57 subfiles exclusively.

Define receiver subset $\Phi\subseteq [3]$ and transmitter subset $\Psi \subseteq [3]$. Then, denote $W_{ir_\Phi t_\Psi}$ as the subfile of $W_i$ cached in receivers in $\Phi$ and transmitters in $\Psi$. For example, $W_{ir_{1}t_{1}}$ is the subfile cached in receiver 1 and transmitter 1,  $W_{ir_{\emptyset}t_{123}}$ is the subfile cached in none of the three receivers but in three transmitters. Similarly, we denote $W_{ir_\Phi}$ as the collection of the subfiles in file $W_i$ that are cached in receivers in $\Phi$, i.e. $W_{ir_\Phi}=\bigcup\limits_{\Psi}W_{ir_\Phi t_\Psi}$. We further assume that the subfiles that are cached in the same number of transmitters and the same number of receivers have the same length. Due to the symmetry of all the nodes as well as the independency of all files, this assumption is valid. Thus, we assume the size of $W_{ir_\Phi t_\Psi}$ is $a_{|\Phi||\Psi|}F$, where $|\Psi|$ and $|\Phi|$ are the cardinalities of $\Psi$ and $\Phi$, respectively, and $a_{|\Phi||\Psi|}$ is the file splitting ratio to be optimized later. For example, the size of $W_{ir_{1}t_{1}}$ is $a_{11}F$ and the size of $W_{ir_{\emptyset}t_{123}}$ is $a_{03}F$. Here, the file splitting ratios $\{a_{|\Phi||\Psi|}\}$ should satisfy the following constraints:
\begin{align}
&a_{30}+3a_{31}+3a_{32}+a_{33}+9a_{21}+9a_{22}+3a_{23}+9a_{11}\notag\\ &+9a_{12}+3a_{13}+3a_{01}+3a_{02}+a_{03}=1, \label{eqn:total cache}\\
&a_{30}+3a_{31}+3a_{32}+a_{33}+6a_{21}+6a_{22}+2a_{23}+3a_{11}\notag\\ &+3a_{12}+a_{13}\le\mu_R, \label{eqn:receiver cache}\\
&a_{31}+2a_{32}+a_{33}+3a_{21}+6a_{22}+3a_{23}+3a_{11}+6a_{12}\notag\\ &+3a_{13}+a_{01}+2a_{02}+a_{03}\le\mu_T. \label{eqn:transmitter cache}
\end{align}

Constraint \eqref{eqn:total cache} comes from the constraint of file size. The multiplier of each splitting ratio $a_{|\Phi||\Psi|}$ in \eqref{eqn:total cache} indicates the number of subfiles that have the same length of $a_{|\Phi||\Psi|}f$. For instance, the number of subfiles with length $a_{21}F$ is nine since there are $\binom{3}{2}\binom{3}{1}=9$ cache states to cache the subfile in two out of the three receivers and one out of the three transmitters. Constraints \eqref{eqn:receiver cache} and \eqref{eqn:transmitter cache} come from the receiver and transmitter cache size limit, respectively. Similar arguments used in \eqref{eqn:total cache} can be applied here to determine the multipliers.

\subsection{File Delivery}
Without loss of generality, we assume that receivers $1,2,3$ desire files $W_1$, $W_2$, $W_3$, respectively. Specifically, receiver $j$ ($j\in[3]$) desires subfiles $W_{jr_\emptyset}$, $W_{jr_{kl}}$, and $W_{jr_{k}}$ that are not cached in its local cache, where $k,l\neq j$. We divide these subfiles into three groups and present the delivery scheme for each group individually.

\subsubsection{Delivery of Subfiles Cached in Two Receivers}
\begin{figure}[tbp]
\begin{centering}
\includegraphics[scale=0.22]{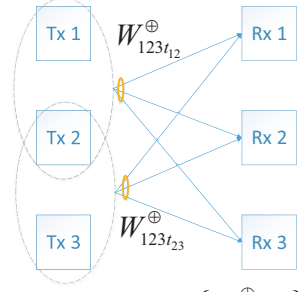}
\vspace{-0.2cm}
\caption{Message flow pattern of $\{W^\oplus_{123 t_\Psi}\}_{|\Psi|=2}$. Only $\Psi=\{1,2\}$ and $\{2,3\}$ are shown. Subfiles are listed beside the channel which carries them. Dashed circle denotes that the transmitters inside it cooperate with each other in the delivery phase. Solid circle denotes that the channels inside it carry the same subfile.}\label{message a2}
\vspace{-0.6cm}
\end{centering}
\end{figure}

Consider the delivery of subfiles $\{W_{jr_{kl}}\}_{k,l\neq j}$ needed by receiver $j$ ($j\in[3]$). Since the subfiles desired by each receiver are cached in the other two receivers, coded multicasting opportunities can be exploited. In specific, consider subfiles $W_{1r_{23}t_\Psi}$, $W_{2r_{13}t_\Psi}$, and $W_{3r_{12}t_\Psi}$ for any transmitter subset $\Psi$. Transmitters in each subset $\Psi$ can generate a new message $W^\oplus_{123t_\Psi}\triangleq W_{1r_{23}t_\Psi}\oplus W_{2r_{13}t_\Psi}\oplus W_{3r_{12}t_\Psi}$ needed by all three receivers, where $\oplus$ denotes the bit-wise XOR. To illustrate the delivery scheme, we take the set of messages with $|\Psi|=2$ for example. The message flow pattern is shown in Fig.~\ref{message a2}. We adopt time division multiple access (TDMA) technique so that all the $\binom{3}{2}$ possible transmitter cooperation sets take turns to transmit. In specific, we divide the transmission time into 3 time slots. In each time slot, we select one transmitter subset $\Psi$ (e.g. $\Psi=\{1,2\}$) and let transmitters in this subset to cooperatively transmit $W^\oplus_{123t_\Psi}$ to all three receivers. The network topology in each slot now becomes a broadcast channel with common information only, which we refer to as \textit{multicast channel}. The maximum sum DoF of the multicast channel is 1, no matter the transmitters cooperate or not. The converse can be proved easily using cut-set bound on each receiver. Thus, the delivery time $T=\frac{3a_{22}F}{\log P}$ is achieved. In the general case with $|\Psi|=i$, we also use the TDMA technique so that all the $\binom{3}{i}$ possible transmitter cooperation sets take turns to transmit. The delivery time is $T=\frac{\binom{3}{i}a_{2i}F}{\log P}$. Thus, the total FDT of subfiles $\{W_{jr_{kl}}\}_{k,l\neq j}$ is $\tau=\frac{1}{3}\sum_{i=1}^{3}\binom{3}{i}a_{2i}$.

\subsubsection{Delivery of Subfiles Cached in One Receiver}
Consider the delivery of subfiles $\{W_{jr_k}\}_{k\neq j}$ needed by receiver $j$ ($j\in[3]$). Since each subfile requested by one receiver is already cached in another receiver, coded multicasting gain can be exploited again. In specific, transmitters in each subset $\Psi$ can generate a new message $W^\oplus_{jkt_\Psi}\triangleq W_{jr_kt_\Psi}\oplus W_{kr_jt_\Psi}$ needed by receivers $j$ and $k$, where $j,k\in[3],j<k$.

\begin{figure*}[tbp]
\begin{minipage}[t]{0.32\linewidth}
\centering
\includegraphics[scale=0.22]{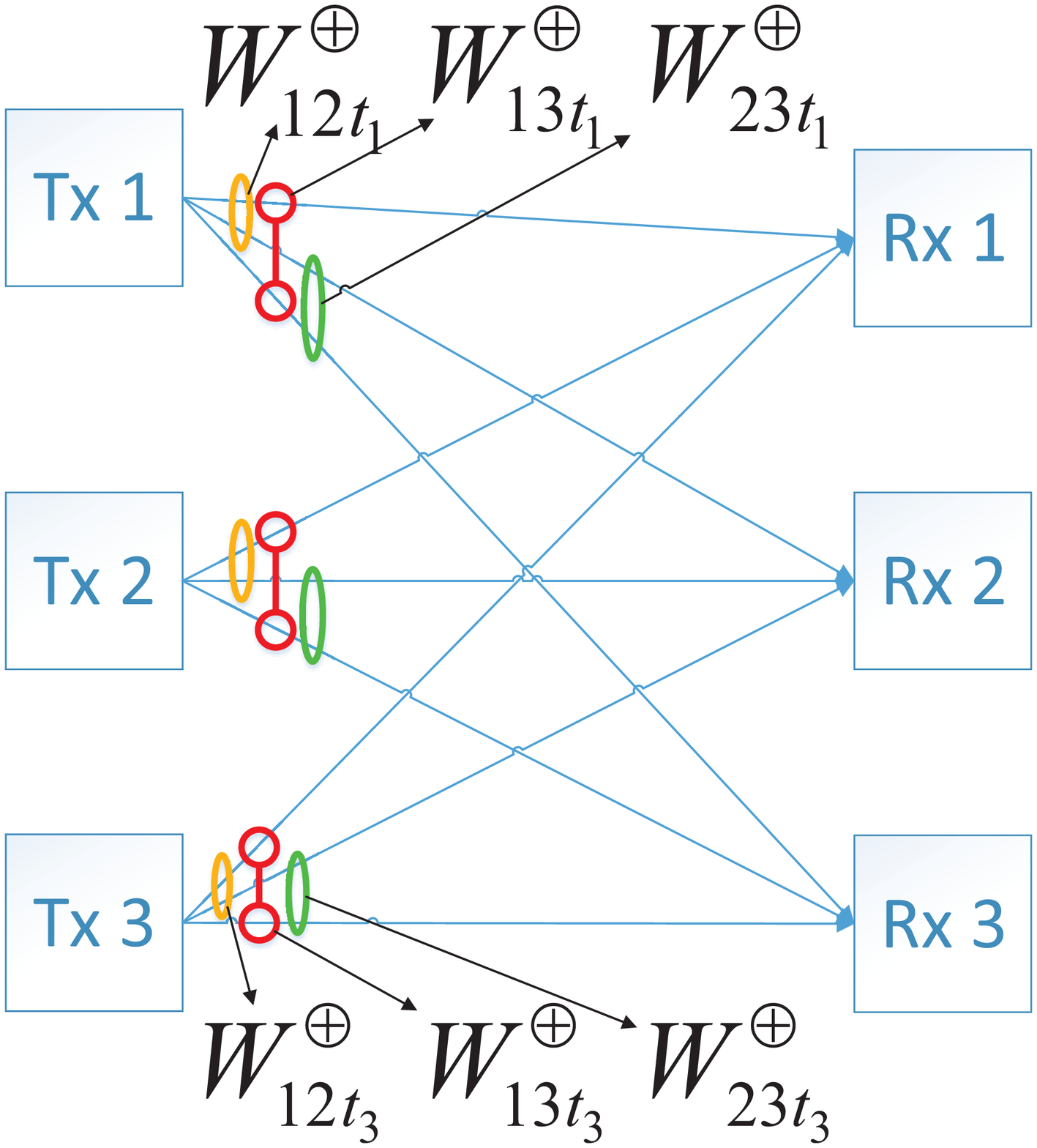}
\vspace{-0.2cm}
\subcaption{}\label{message a11}
\end{minipage}
\begin{minipage}[t]{0.32\linewidth}
\centering
\includegraphics[scale=0.22]{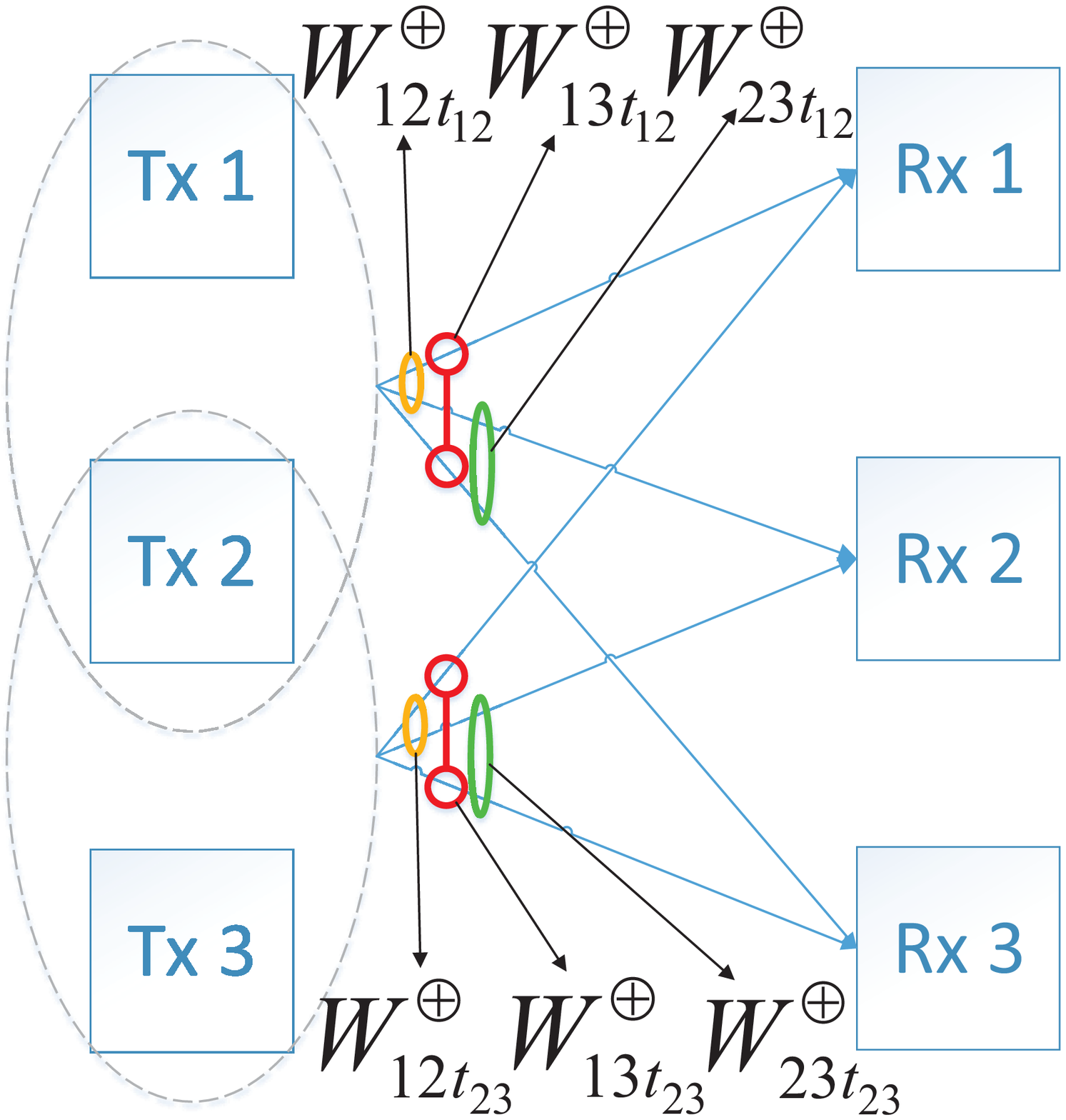}
\vspace{-0.2cm}
\subcaption{}\label{message a12}
\end{minipage}
\begin{minipage}[t]{0.32\linewidth}
\centering
\includegraphics[scale=0.22]{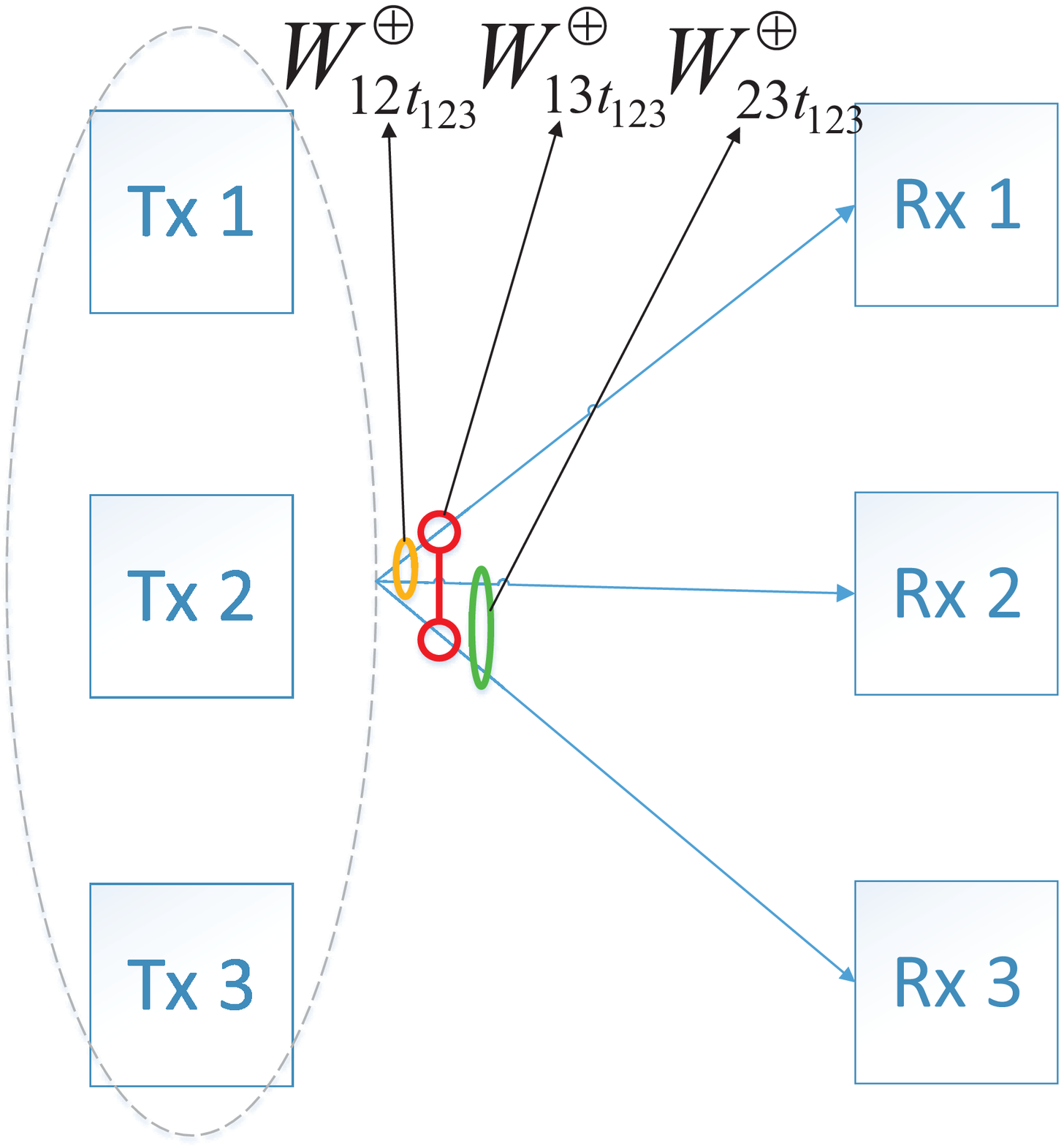}
\vspace{-0.2cm}
\subcaption{}\label{message a13}
\end{minipage}
\vspace{-0.2cm}
\caption{Message flow pattern of (a) $\{W^\oplus_{jkt_\Psi}\}_{|\Psi|=1}$, (b) $\{W^\oplus_{jkt_\Psi}\}_{|\Psi|=2}$, only $\Psi=\{1,2\}$ and $\{2,3\}$ are shown, (c) $\{W^\oplus_{jkt_\Psi}\}_{|\Psi|=3}$.}\label{message a1}
\vspace{-0.6cm}
\end{figure*}

We first consider the delivery of messages $\{W^\oplus_{jkt_\Psi}\}_{j<k}$ with $|\Psi|=1$. The message flow pattern is shown in Fig.~\ref{message a11}, and the network topology can be seen as the hybrid X-multicast channel. Lemma 1 below presents the sum DoF of this channel. The proof is based on interference alignment and given in \cite[Appendix B]{myarkiv}.

\textbf{Lemma 1}: The achievable sum DoF of the $3\times 3$ hybrid X-multicast channel is $\frac{9}{7}$.

Using Lemma 1 and given that the total amount of bits to deliver is $9a_{11}F$, we obtain $\tau=\frac{7a_{11}}{3}$.

Next, we consider the delivery of messages $\{W^\oplus_{jkt_\Psi}\}_{j<k}$ with $|\Psi|\ge2$. The message flow patterns for $|\Psi|=2$ and $|\Psi|=3$ are shown in Fig.~\ref{message a12} and \ref{message a13}, where the network topologies can be seen as the partially and fully cooperative hybrid X-multicast channel, respectively. Lemma 2 below presents the sum DoF of this channel. Its proof is based on interference neutralization and given in \cite[Appendix C]{myarkiv}.

\textbf{Lemma 2}: The achievable sum DoF of the $3 \times 3$ partially or fully cooperative hybrid X-multicast channel is $\frac{3}{2}$.

As such, the total FDT of subfiles $\{W_{jr_kt_\Psi}\}_{j,k\in[3],j<k}$ for $\Psi$'s with $|\Psi|=2,3$ is $\tau=\frac{6a_{12}+2a_{13}}{3}$.

\subsubsection{Delivery of Subfiles Cached in None Of Receivers}
Consider the delivery of subfiles $\{W_{jr_{\emptyset}}\}$ needed by receiver $j$ ($j\in[3]$). Each $W_{jr_{\emptyset}}$ further consists of subfiles $W_{jr_{\emptyset}t_\Psi}$ for all transmitter subsets $\Psi$'s with $|\Psi|=1,2,3$. The message flow patterns of $\{W_{jr_{\emptyset}t_\Psi}\}_{|\Psi|=3}$, $\{W_{jr_{\emptyset}t_\Psi}\}_{|\Psi|=2}$ and $\{W_{jr_{\emptyset}t_\Psi}\}_{|\Psi|=1}$ correspond to the patterns in \cite{upperbound} when $\mu_T=1,\frac{2}{3},\frac{1}{3}$, respectively. In \cite{upperbound}, the message flow patterns of $\{W_{jr_{\emptyset}t_\Psi}\}_{|\Psi|=3}$, $\{W_{jr_{\emptyset}t_\Psi}\}_{|\Psi|=2}$, and $\{W_{jr_{\emptyset}t_\Psi}\}_{|\Psi|=1}$ form a MISO broadcast channel, a partially cooperative X channel, and an X channel, respectively. Thus, the delivery time of subfiles $\{W_{jr_\emptyset t_\Psi}\}_{j\in[3]}$ for all $\Psi$'s is $T=\frac{3a_{03}F}{3\log P}+\frac{9a_{02}F}{18\log P/7}+\frac{9a_{01}F}{9\log P/5}$ and its corresponding FDT is $\tau=\frac{a_{03}}{3}+\frac{7a_{02}}{6}+\frac{5a_{01}}{3}$.

\section{Caching Optimization}
Combining all the FDTs obtained in Section IV-B, we obtain the total FDT in the delivery phase as
\begin{align}
\tau=\frac{1}{3}&(5a_{01}+\frac{7}{2}a_{02}+a_{03}+3a_{21}+3a_{22}+a_{23}\notag\\&+7a_{11}+6a_{12}+2a_{13}). \label{original tau}
\end{align}
Our goal is to minimize the FDT subject to the file slitting ratio constraints \eqref{eqn:total cache}\eqref{eqn:receiver cache}\eqref{eqn:transmitter cache}. This is formulated as:
\begin{align}
&\min\;\tau(\mu_R,\mu_T)\rm \label{eqn original problem}\\&s.t.\quad \eqref{eqn:total cache}\eqref{eqn:receiver cache}\eqref{eqn:transmitter cache},\notag
\end{align}
which is a standard linear programming problem. Using linear equation substitution and other manipulations, we can obtain the optimal solutions in closed form as follows. Here, all the regions are defined in Theorem 1.


\textit{Region $\mathcal{R}_1$}:\quad The optimal FDT is $\tau^*=\frac{1}{3}-\frac{\mu_R}{3}$. The optimal splitting ratios are not unique but must satisfy that $a^*_{11}=a^*_{01}=a^*_{02}=0$ and that the equality in \eqref{eqn:receiver cache} holds. One feasible solution is $a^*_{30}=\mu_R$, $a^*_{03}=1-\mu_R$ and other ratios are 0.

\textit{Region $\mathcal{R}_2$}:\quad The optimal FDT is $\tau^*=\frac{4}{9}-\frac{4\mu_R}{9}-\frac{\mu_T}{9}$. The optimal splitting ratios are not unique but must satisfy
\begin{align}
\left\{
\begin{array}{ll}
a^*_{01}=a^*_{02}=a^*_{31}=a^*_{32}=a^*_{33}=a^*_{22}=a^*_{23}=a^*_{13}=0\\
a^*_{11}=\frac{1}{3}-\frac{\mu_R}{3}-\frac{\mu_T}{3}\\
a^*_{30}+6a^*_{21}+3a^*_{12}=2\mu_R+\mu_T-1\\
3a^*_{21}+6a^*_{12}+a^*_{03}=\mu_R+2\mu_T-1
\end{array}.
\right.\notag
\end{align}
One feasible solution is $a_{11}^*=\frac{1}{3}-\frac{\mu_R}{3}-\frac{\mu_T}{3}$, $a_{30}^*=2\mu_R+\mu_T-1$, $a_{03}^*=\mu_R+2\mu_T-1$ and other ratios are 0.
%

\textit{Region $\mathcal{R}_3$}:\quad The optimal FDT is $\tau^*=\frac{1}{2}-\frac{5}{9}\mu_R-\frac{1}{6}\mu_T$. The optimal splitting ratios are unique and given by $a^*_{11}=\frac{\mu_R}{3}$, $a^*_{02}=1-2\mu_R-\mu_T$, $a^*_{03}=3\mu_R+3\mu_T-2$ and other ratios being 0.



\textit{Region $\mathcal{R}_4$}:\quad The optimal FDT is $\tau^*=\frac{13}{18}-\frac{8}{9}\mu_R-\frac{1}{2}\mu_T$. The optimal splitting ratios are unique and given by $a^*_{11}=\frac{\mu_R}{3}$, $a^*_{01}=\frac{2}{3}-\mu_R-\mu_T$, $a^*_{02}=\mu_T-\frac{1}{3}$ and other ratios being 0.



\textit{Region $\mathcal{R}_5$}:\quad The optimal FDT is $\tau^*=\frac{8}{9}-\frac{8}{9}\mu_R-\mu_T$. The optimal splitting ratios are unique and given by $a^*_{11}=\frac{\mu_R}{3}+\mu_T-\frac{1}{3}$, $a^*_{01}=1-\mu_R-2\mu_T$, $a^*_{30}=1-3\mu_T$ and other ratios being 0.

Summarizing all the results above, we finish proof of Theorem 1.

\textbf{Remark 4}: In $\mathcal{R}_1$ and $\mathcal{R}_2$, the multiple choices of file splitting ratios from caching optimization offer freedom to choose appropriate caching and delivery scheme in practical systems according to different limitations, such as file splitting constraints.

\textbf{Remark 5}: In the proposed caching strategy, the local caching gain, transmitter cooperation gain, and coded multicasting gain are exploited opportunistically in different cache size regions. These gains are reflected by the file splitting ratios of the corresponding cache states. In $\mathcal{R}_1$, local caching gain and cooperation gain are exploited, because the feasible solution is $a^*_{30}=\mu_R$, $a^*_{03}=1-\mu_R$. We do not need to use up the total cache storage at transmitters. In $\mathcal{R}_2$, $\mathcal{R}_3$, and $\mathcal{R}_4$, all the three gains are exploited since their optimal solutions all satisfy $a^*_{11}>0$ and $a^*_{02}+a^*_{03}>0$. In $\mathcal{R}_3$ and $\mathcal{R}_4$, there do not exist two receivers which cache the same content. Instead, each receiver uses up its total cache size to cache the content already cached in only one transmitter, i.e. $W_{ir_jt_p}$, to fully exploit coded multicasting gain. In $\mathcal{R}_5$, only local caching gain and coded multicasting gain are exploited and no transmitter cooperation can be exploited since the optimal solution satisfies $a^*_{mn}=0$, $\forall n\ge2$. This is due to that the transmiter cache size is approaching its lower limit $\mu_T\ge\frac{1}{3}(1-\mu_R)$ in \eqref{eqn boundary}.

\textbf{Remark 6}: Although the similar caching problem is considered in \cite{bothcache}, their performance metric, caching scheme, and conclusion are significantly different from ours. First, we adopt the FDT as the performance metric, while \cite{bothcache} used the standard DoF. From Remark 2, FDT reflects not only the load reduction due to receiver cache but also the DoF enhancement due to transmitter cache, while the DoF alone cannot reflect the former one. Also, at each given $(\mu_R, \mu_T)$, the file splitting ratios in \cite{bothcache} are pre-determined, while our file splitting ratios are obtained by solving a linear programming problem and thus are probably optimal under the given caching strategy. Another difference is that the transmission scheme in \cite{bothcache} is restricted to one-shot linear processing, while we allow interference alignment which may require infinite symbol extension.

\bibliographystyle{IEEEtran}
\bibliography{IEEEabrv,journal}

\end{document}